\documentclass[journal]{IEEEtran}
\usepackage{graphicx}
\usepackage{multirow}
\usepackage[bookmarks=false]{hyperref}
\hypersetup{
    bookmarks    = false,            
    unicode      = false,            
    pdftoolbar   = false,             
    pdfmenubar   = false,             
}
\begin{document}
%
\title{Performance Evaluation of Widely\\
used Portknoking Algorithms}

\author{\IEEEauthorblockN{Z. A. Khan$^{\S}$, N. Javaid, M. H. Arshad, A. Bibi, B. Qasim\\
                Department of Electrical Engineering, COMSATS\\ Institute of
                Information Technology, Islamabad, Pakistan. \\
                $^{\S}$Faculty of Engineering, Dalhousie University, Halifax, Canada.}}

\maketitle

\begin{abstract}
Port knocking is a technique by which only a single packet or special sequence will permit the firewall to open a port on a machine where all ports are blocked by default. It is a passive authorization technique which offers firewall-level authentication to ensure authorized access to potentially vulnerable network services. In this paper, we present performance evaluation and analytical comparison of three widely used port knocking (PK) algorithms, Aldaba, FWKNOP and SIG-2. Comparative analysis is based upon ten selected parameters; Platforms (Supported OS), Implementation (PK, SPA or both), Protocols (UDP, TCP, ICMP), Out of Order packet delivery, NAT (Network Address Translation), Encryption Algorithms, Root privileges (For installation and operation), Weak Passwords, Replay Attacks and IPv6 compatibility. Based upon these parameters, relative performance score has been given to each algorithm. Finally, we deduce that FWKNOP due to compatibility with windows client is the most efficient among chosen PK implementations.
\end{abstract}

\begin{IEEEkeywords}
Port Knocking; Firewall knock operator; User Datagram Protocol; Authenticating User; Internet Protocol ;Single Packet Authorization;Network Address Translator; Man In the Middle; Secure Shell; Transmission Control Protocol
\end{IEEEkeywords}

\section{Introduction}

\IEEEPARstart{I}{NTERNET} became part of our life fourty-one years ago. From the beginning we know that Internet is acting as a hostile place. So it is important to keep a check on unauthorized intrusion and other harmful invaders. The importance of securing the hostile world of internet has increased now, because there have not been such deadly risks before. Reason of this increased security risk is the introduction of Internet. The only secured system is the one that has no connection with the outside world. But internet is the other name of connection, so it is impossible to have no connection with the world and have Internet at the same time.  Only thing that can be done is to limit the number of people and set of instructions accessible to the computer.

Many security schemes have been developed to fight against the attacks and risks but the attackers have improved within same manners. One way to limit access to selected users is by using an authentication method, but this is not a perfect solution. User first to prove its identity and then access is granted to that user upon verification of authenticity. Many large and complicated systems have suffered from flaws in their authentication mechanisms, which make secure systems vulnerable to the attackers. One usual method of limiting the hosts is to use firewall.

\begin{figure}[h]
\begin{center}
\includegraphics[scale=0.40]{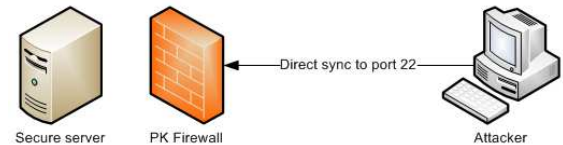}
\caption{All well-known ports of secure server are closed by PK firewall so when an attacker try to directly get connected to port 22, the firewall will simply drop its packet and will not allow attacker to access any secure port directly}
\end{center}
\end{figure}

Firewall selectively accepts and rejects network packets by considering their source address and other important characteristics. Some dangerous attackers are capable enough to hide the source of packets sent by them. Users having unpredictable IP can also easily pass through firewall. So firewall is also not a complete solution as well.

\begin{figure}[h]
\begin{center}
\includegraphics[scale=0.40]{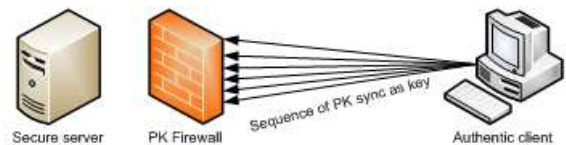}
\caption{An authentic client who knows the predefined sequence of knocks which will act as a key will send tcp sync packets to those pre-defines sequence of ports}
\end{center}
\end{figure}

Port knocking is a kind of security mechanism installed over firewall of secure computer systems. Basically what port knocking does is that it provides with another security layer over the security we already have. PK close all ports of the system on which it is implemented, It also works on the principle of least privileged as it blocks all the unauthorized users at first and we can say that there is a visible enhancement in the security to that of a system with no port knocking mechanism, PK scenario is explained through figures from 1-a to 1-d.

The steps involved in PK authentication can be clearly understood through the flowchart in figure 2. The flow chart provides a detailed step wise complete procedure of a general PK authentication mechanism. Since the advent of PK authentication scheme a lot of PK algorithms have been presented with different characteristics. So what we have done is carried out a performance evaluation and analytical comparison of three widely used PK algorithms against ten different parameters.

\begin{figure}[h]
\begin{center}
\includegraphics[scale=0.40]{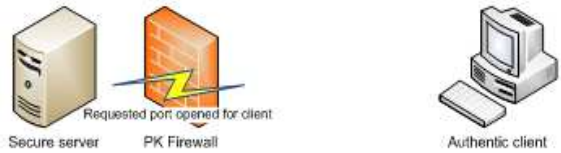}
\caption{PK demon installed on secure server firewall will silently watch those packets and if theses port knocks found to be in correct pre-defined order the client will be considered as authentic client, and PK demon will open clients requested port.}
\end{center}
\end{figure}

\begin{figure}[h]
\begin{center}
\includegraphics[scale=0.40]{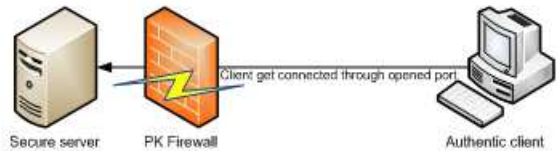}
\caption{Authentic client after successful authentication will get connected to the secure server through one of the well-known port opened by the PK demon}
\end{center}
\end{figure}

\begin{figure*}[t]
\begin{center}
\includegraphics[scale=0.60]{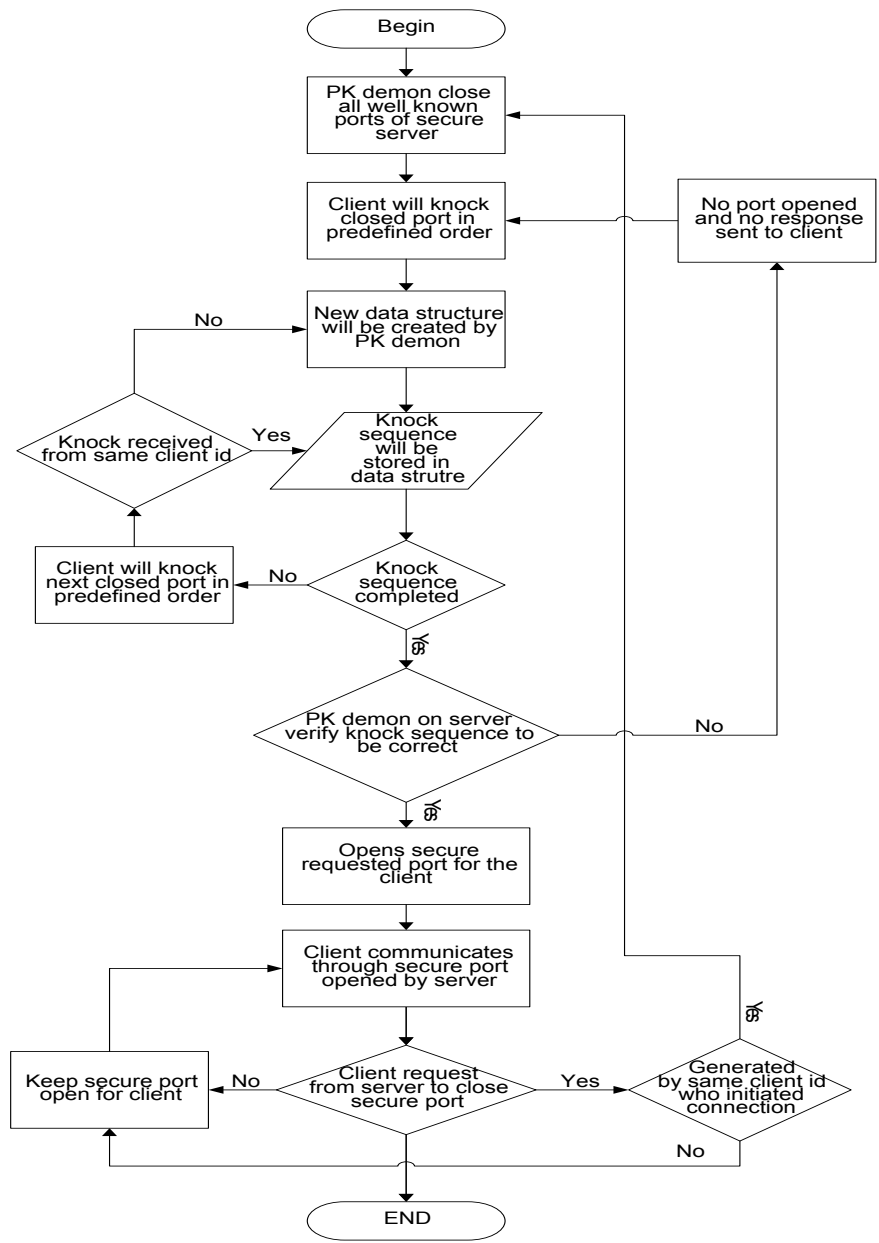}
\caption{Port Knocking Flowchart}
\end{center}
\end{figure*}

In section II related work and motivation is described, section III contains description of all ten parameters on basis of which performance evaluation will be done, after that limits of scores will be defined and a performance evaluation will be carried out assigning scores to a particular algorithm against ten parameters according to its performance and compatibility with that parameter. Then separate graphs will be presented for all three algorithms which represents their scores against ten parameters, and then combining all these three graphs an overall comparison graphs is plotted to comparatively demonstrate their scores and distinguish the best one out of these three, in the same section we have also presented some unique features of all three PK algorithms. Then finally in section IV we concluded this analytical comparison and defined the best algorithm against ten parameters.

\section{Related Work and Motivation}
	
Hussein Al-Bahadili  \cite{1}; develops and evaluate parameters of newly built PK implementation referred with the name of hybrid, which have the capability to defeat previous knocking techniques. This new technique uses concepts of PK, mutual authentication and steganography.

RenniedeGraaf, Improved Port Knocking with Strong Authentication \cite{2}; studies existing PK implementation, improves existing PK techniques, builds a new PK technique which is refers to as novel port knocking technique.

Authors in  \cite{3}; presents improvements in existing PK and SPA techniques like using one time password method using cellular networks such as CDMA, GSM to enhance security, defines protection against dictionary attacks.

Muhammad Tariq et al. Associating the Authentication and Connection Establishment Phases in Passive Authorization Techniques, Proceedings of the World Congress on Engineering 2008 , London, U.K  \cite{4}; define weaknesses like lack of link between establishment of a TCP connection, highlight authentication process, presents another novel PK technique, simulation carried out to evaluate algorithms on the basis of overhead calculations.

KonstantinosXynos and Andrew Blyth  \cite{5}; propose an idea of implementing port knocking technique over a gateway authentication layer or gateway authentication program or network service program instead of firewall, eliminate problems with firewalls and reduce brute force attacks.

Ben Maddock  \cite{6}; defines portknocking and its benefits in detail, elaborates features of existing portknocking techniques, finally future offer exploration and PK conclusion.

Dawn Isabel, Port Knocking: Beyond the Basics  \cite{7}; provides three solutions for two basic problems with static PK i.e. detection and replay, propose solutions of dynamic knocks, covert knocks, and one time knocks; implementing these solutions over four PK techniques.

Sebastien Jeanquier  \cite{7} in his MS thesis, "An Analysis of Port Knocking and Single Packet Authorization"; analyzes PK and SPA as network security mechanisms, performs compatibility as firewall authentication schemes and discusses drawbacks and outcomes in current PK implementations, critical evaluation of FWKNOP, outlining its outcomes and suggesting some remedies.

Work done by Sabastien is of great regard as it provides evaluation of a single PK over several parameters but there can also be a research of several widely used PK implementations against different parameters so that a new person in this field can come to know that which implementation is the best, so we have done this work in this research paper by analytically evaluating several widely used PK implementations on ten parameters. We have presented our data with the help of graphs to provide an even better view to the reviewer.

\section{Performance Evaluation PK Algorithms}

In this paper we have evaluated the performance of three PK algorithms under different scenarios and parameters. Then we presented their performance comparison. For this purpose we have selected   following ten performance parameters.
Parameters:

	Platforms (Supported OS)

	Implementation (PK, SPA or both)

	Protocols (UDP, TCP, ICMP)

	Out of Order packet delivery

	NAT (Network Address Translation)

	Encryption Algorithms

	Root privileges (For installation and operation)

	Weak Passwords

	Replay Attacks

 	IPv6 compatibility

We have plotted graphs of each algorithm against these parameters. The range of performance score is 0-100. It means that the better is the performance of the algorithm against a parameter the better is the score assigned to that algorithm. Which means that a score of 100 will be awarded to that algorithm which will  fully supports the aspects of the given parameter and has solution to all the issues related with that metric. Similarly if algorithm is not robust enough against that parameter then it will get a relatively less score.

\begin{figure}[h]
\begin{center}
\includegraphics[scale=0.40]{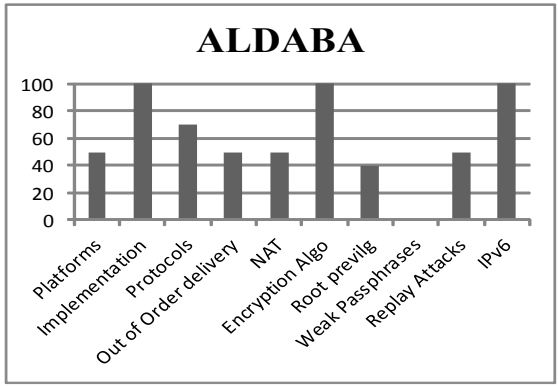}
\caption{Performance graph of Aldaba}
\end{center}
\end{figure}

Platforms means the OS which are supported, in FWKNOP the client can supports both Windows and UNIX based versions but it can only have a UNIX based server hence it has given score of 80, while Aldaba scores a 50 due to the presence of only UNIX based client and server, sig-2 has maximum score due to both Windows and UNIX based client and server.

Implementation stands for the Port Knocking scheme use either PK or SPA or both. In this case Aldaba scores 100 due to support with both PK and SPA, on the other hand FWKNOP and SIG-2 both   scores a 50 because they only use one implementation.

\begin{figure}[h]
\begin{center}
\includegraphics[scale=0.40]{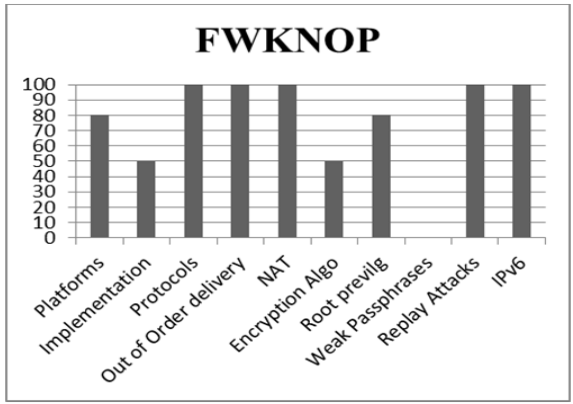}
\caption{Performance graph of FWKNOP}
\end{center}
\end{figure}

FWKNOP supports three protocols namely UDP, TCP and ICMP so it has been awarded maximum score i.e 100, on the other hand Aldaba supports UDP and TCP so it has been awarded 70 points whereas SIG-2 supports only TCP so it has been given 50 points.

The problem of Out of Order packet delivery is inherent in contemporary networks, in FWKNOP this is not an issue because it uses SPA which comprises of only a single packet hence it has maximum score in this regard. In SIG-2, there is no solution for this problem and so, no score, while Aldaba handles it by using Sequence Numbers, so it has maximum score.

NAT is widely used in present day networks and the problem is that the client has to know its public IP address before implementing PK as it can't include it private IP address in the authorization packet. FWKNOP automatically obtains the public IP address, so, it earns maximum score. Aldaba resolves this issue by letting the client specify its public IP address, as this puts the client in some misery so Aldaba has score of 50. SIG-2 has no solution to this issue so no score for this.

\begin{figure}[h]
\begin{center}
\includegraphics[scale=0.40]{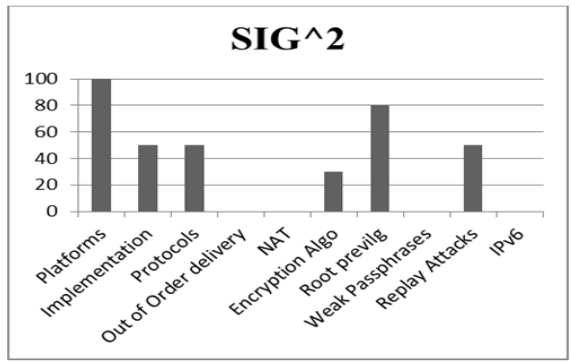}
\caption{Performance graph of SIG-2}
\end{center}
\end{figure}

Encryption is a vital part of PK and the more encryption algorithms a PK implementation supports the better it is. Hence Aldaba has the maximum score because it supports 5 encryption algorithms compared with 2 in FWKNOP and only 1 in SIG-2 who has score of 50 and 30 respectively.

In FWKNOP and SIG-2 root privileges are required for installation only, which gives it 80 score. Aldaba on the other hand requires it for both installation and operation hence getting only a score of 50.

If the passphrases which are used to encrypt and decrypt the PK packets or SPA packet are weak [7] i.e vulnerable to dictionary and Brute Force attacks then an attacker in the Man in the Middle Position can easily capture the authorization packet and can obtain the passphrase, once he has got the passphrase he can decrypt the packet and use the information for crafting his own packet. Unfortunately none of these three PK implementations has solution to this problem. Hence no score is given to any of the three algorithms.

An eavesdropper inside network [7] is who has ability to watch traffic between client and server. It means that attacker can replay the authorization packet on its way from client to server and replay it at a later time. In this way attacker can gain access to the server by replacing the IP address as server has no method to know that the received packet is from a valid client or it is just a replay packet by an attacker. FWKNOP solves this problem by including a timestamp inside authorization packet which is accurate up to minutes and also it includes some random data. The presence of timestamp and random data ensures that the received packet is fresh packet and not an old replayed packet so it gives maximum score to FWKNOP due to through resolution of this issue. In Aldaba this issue is resolved by including the IP address of the client inside authorization packet but it is not a complete solution because an attacker can still modify the IP address if he knows the passphrase, so we give Aldaba a 50 score for this parameter. SIG-2 also uses timestamp but it doesn't have the feature of random data so it also scores 50.

Both Aldaba and FWKNOP are fully compatible with IPv6 which gives them a maximum score whereas SIG-2 is not compatible with IPv6 so no score for it.

\begin{figure}[h]
\begin{center}
\includegraphics[scale=0.40]{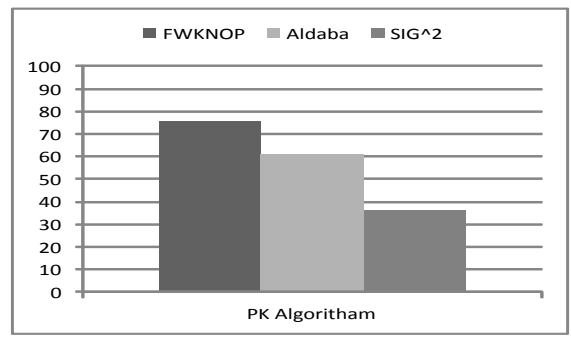}
\caption{Overall Comparison}
\end{center}
\end{figure}

\subsection{FWKNOP Unique Features}

Besides these parameters there are also some features which are unique to these PK implementations. In this section we will also discuss them.

FWKNOP has features like Port randomization support for target port of SPA packets and the port over which secondary connection is made using iptables. Later to access granted to local sockets on the system running FWKNOP and to forward such connections to internal services.

FWKNOP also has comprehensive test suite which allows series of tests which are designed to validate both the client and server pieces either they are installed properly or not. Tests sniff SPA packets throughout the interface of local loopback, this builds brief firewall rules which are verifies against the particular access based on some testing configuration, later analyzing output of fwknop client and server for predictable outcomes for each test. This result can be utilized for sake of communication with the third parties investigation.

Multiple users at the same time are also supported by FWKNOP, every user is assigned its own symmetric or asymmetric encryption key through /etc/fwknop/access.conf file.

Outfits versioned protocol of SPA, and makes it convenient to extend protocol offer for another SPA message and at the same time maintains backwards compatibility with other FWKNOP clients at the same time.
Fwknop also implies execution of shell commands on basis of effective SPA packets.

\subsection{Aldaba Unique Features}

Application which produces packets at its own should be careful in the creation of process. Latest protocols must support some authentic checksums, special byte orders, particular values, etc. If packets do not follow these principles, routers, firewalls and other such network devices can have problems resulting in discard of packets before reaching to their host.
Multiple knocking attempts should be supported for systems that have two or more than two users so that they can make themselves able to handle multiple listening at a time. In a case where system is having some deficiencies in its design, we can have problem when two clients send different knocks at same time can interfere with knocks of each other which will result as a DOS (denial of service) for both of clients.

Knock with originating IP address should associate any PK implementation. This can be done quite easily, so that is why most of the PK systems use it. Actual problem is that an attacker can detect start of knock sequence, and then he can fake his own packets with any kind of random data by deceiving the client's IP address and sending this information to knocking server with in a valid knock causing server to evaluate incorrect data resulting in discard of knock and denial of service for client.

This problem is not present in Aldaba in the case where authentication protocol is SPA due to the reason that only single packet is involved in such process. Port Knocking is vulnerable to such an attack. Till now Aldaba does not have any proper solution for this problem so still there is a chance that a client will suffer DOS attack in a case when an attacker is able to guess knocking sequence and detect the start of a knocking attempt made by client.

Port knocking causes additional load while listening to the incoming packets due to process being done by knocking server where ports scanning is carried out.

Also, whenever start of a knocking attempt is detected, then new data structures will be generated in order to handle them.
Moreover, Aldaba also keeps complete record of all knocking attempts that are under process or have to be processed. New data entry is created whenever start of a knock is detected, if an attacker comes to know about the ports that forms knocking sequence, it will be able to create and send different source IP addresses of multiple packets. In such a case knocking server treat this situation as multiple clients are trying to send a knock so a node will be created for each different IP. If attacker will keep on sending false knocks then a time will come when system will eventually run out of memory and the ultimate result will be crash the server.

\subsection{SIG-2 unique features}

Sig-2 does not contain any such unique features other than those which we have already mentioned above.

\section{Conclusion}

After the performance evaluation it is concluded that SIG-2 Port Knocking is a backward and weak implementation as we can clearly see through graphs. FWKNOP and Aldaba Port Knocking are good implementation with nearly same features. Ability of FWKNOP to use windows client as well gives it a slight edge over Aldaba port Knocking.

\ifCLASSOPTIONcaptionsoff
  \newpage
\fi

\end{document}